\documentclass[]{aastex631}

\usepackage{natbib}
\usepackage{color}
\usepackage{xspace}
\usepackage{amsmath}

\newcommand{\PAR}{\mathrm{P}}
\newcommand{\bPAR}{\mathbf{P}}

\newcommand{\mil}{\mathit{l}}

\shorttitle{}
\shortauthors{}
\graphicspath{{./}{./}}

\begin{document}

\title{Stochastic Averaging of Radiative Transfer Coefficients for Relativistic Electrons}

\correspondingauthor{Monika Mo{\'s}cibrodzka}
\email{m.moscibrodzka@astro.ru.nl}

\author[0000-0002-4661-6332]{Monika Mo{\'s}cibrodzka}
\affiliation{Department of Astrophysics/IMAPP, Radboud University,P.O. Box
  9010, 6500 GL Nijmegen, The Netherlands}

\author[0000-0001-7451-8935]{Charles F. Gammie}
\affiliation{Astronomy, Physics, NCSA, and ICASU, University of Illinois, Urbana, Illinois 61801, USA}

\begin{abstract}

Synchrotron emissivities, absorptivities, and Faraday rotation and conversion coefficients are needed in modeling a variety of astrophysical sources, including Event Horizon Telescope (EHT)  sources.  We develop a method for estimating transfer coefficients that exploits their linear dependence on the electron distribution function, decomposing the distribution function into a sum of parts each of whose emissivity can be calculated easily.  We refer to this procedure as stochastic averaging and apply it in two contexts.  First, we use it to estimate the emissivity of an isotropic $\kappa$ distribution function with a high energy cutoff. The resulting coefficients can be evaluated efficiently enough to be used directly in ray-tracing calculations, and we provide an example calculation. Second, we use stochastic averaging to assess the effect of subgrid turbulence on the volume-averaged emissivity and along the way provide a prescription for a turbulent emissivity.  We find that for parameters appropriate to EHT sources turbulence reduces the emissivity slightly.  In the infrared turbulence can dramatically increase the emissivity.   

\end{abstract}


\section{Introduction}

Semi-analytic and numerical models of black hole accretion flows, including general relativistic magnetohydrodynamics simulations, are now routinely used for physical interpretation of observational data \citep[e.g.,][]{paper5:2019,paper5:2022}.  In many cases the dominant radiative process at the frequencies of interest is synchrotron; scattering is typically negligible.  Synchrotron photons are emitted, absorbed, and their polarization is modified by interaction with the plasma. In the radiative transfer equation these interactions are encoded in the transfer coefficients for emission, absorption, and Faraday rotation (rotation of the polarization) and conversion (conversion of circular to linear polarization and vice versa).  The transfer coefficients depend on the electron distribution function $f$.  

Calculation of synchrotron transfer coefficients from $f$ is notoriously difficult \citep{schwinger:1949}.  Methods for numerically evaluating coefficients for isotropic distribution functions now exist \citep{marszewski:2021}, or even some coefficients for anisotropic distribution function \citep{verscharen:2018, galishnikova:2023}. Although the direct numerical evaluation of coefficients is possible, direct evaluation is not practical in numerical radiative transfer calculations.  

Instead simple, easy to evaluate analytic approximations are typically used for a small set of common model distribution functions.  For example, useful expressions exist for the relativistic isotropic thermal (Maxwell-J{\"u}ttner) and non-thermal (power-law or $\kappa$) electron distribution functions \citep[e.g.,][]{shcherbakov:2008,huang:2011,leung:2011,dexter:2016,pandya:2016,pandya:2018,marszewski:2021}. Nevertheless, one might want to evaluate a model for a new or slightly modified distribution function - for example a power law distribution with an imposed exponential cutoff.  In general this would require an expensive numerical re-evaluation and re-fitting of the coefficients.

Here we propose a method for evaluating synchrotron transfer coefficients semi-analytically.  Our method is based on the observation that the coefficients depend linearly on the distribution function $f$, so that, for example, the emissivity $j_\nu(f_1 + f_2) = j_\nu(f_1) + j_\nu(f_2)$.  This observation has been made before in the context of accreting black holes by \citet{mao:2017}.  It implies that if we can write the distribution function as a weighted sum of distribution functions for which accurate but approximate analytic transfer coefficients are known, then we can also evaluate the transfer coefficients as a weighted sum. Here we apply this notion in two contexts.  

First, we develop an accurate, efficient procedure for estimating transfer coefficients for the $\kappa$ distribution function, which has been used in studies of black hole accretion and jets to model electron acceleration processes (e.g., \citealt{davelaar:2018,paper5:2019,davelaar:2023}). The $\kappa$ distribution smoothly connects a quasi-thermal core at low momentum to a power-law tail at high momentum.  It is known to be a good fit to  electron populations in the solar wind \citep{vasyliunas:1968}; for a recent review see \cite{pierrard:2022}.  We also show how our procedure makes it easy to incorporate a cutoff in the distribution function.  

The application to $\kappa$ distributions is inspired by work on ``stochastic averaging''  by \citet{schwadron:2010}, and we adopt this term to describe our method of taking weighted averages of emission coefficients. 

Second, we use stochastic averaging to write down turbulent transfer coefficients, which may be useful in estimating the emergent radiation from numerical simulations.  Large eddy simulations of turbulence are common in astrophysics.  Each resolution element, or zone, represents a region with unresolved turbulent substructure.  The effect of that substructure on the radiation field can be modeled by replacing the emission associated with each zone's temperature, density, velocity, and magnetic field strength by a weighted average of emissivities over a distribution of temperatures, densities, etc.  We provide a general expression for turbulent emissivity that depends on the covariances of the emissivity parameters in a turbulent flow, and estimate the magnitude of the change in emissivities for models of Sgr A*.  

This paper is organized as follows.  Section~\ref{sec:2} describes various distribution functions and develops the necessary notation. Section~\ref{sec:3} applies stochastic averaging to the $\kappa$ distribution function.  Section \ref{sec:4} applies stochastic averaging to turbulent transfer coefficients.  Section \ref{sec:5} summarizes and identifies directions for future research.  

\section{Preliminaries}\label{sec:2}

The polarized radiative transfer equation for the intensities in Stokes parameters $I,Q,U,V$ along a coordinate $s$ is
\begin{equation}
    \frac{d}{ds} 
    \begin{pmatrix} 
    I_\nu \\ Q_\nu \\ U_\nu \\ V_\nu
    \end{pmatrix}
    = 
    \begin{pmatrix} 
    j_{\nu,I} \\ j_{\nu,Q} \\ j_{\nu,U} \\ j_{\nu,V}
    \end{pmatrix}
    -
    \begin{pmatrix}
        \alpha_{\nu, I} & \alpha_{\nu, Q} & \alpha_{\nu, U} & \alpha_{\nu, V} \\
        \alpha_{\nu, Q} & \alpha_{\nu, I} &
        \rho_{\nu, V} & - \rho_{\nu, U} \\
        \alpha_{\nu, U} & -\rho_{\nu, V} & \alpha_{\nu, I} & \rho_{\nu, Q} \\
        \alpha_{\nu, V} & \rho_{\nu, U} & -\rho_{\nu, Q} & \alpha_{\nu, I}         
    \end{pmatrix}
    \begin{pmatrix} 
    I_\nu \\ Q_\nu \\ U_\nu \\ V_\nu
    \end{pmatrix}.
\end{equation}
where we have neglected scattering.  Here $j$ designates an emissivity, $\alpha$ an absorptivity, and $\rho$ a rotativity, i.e. a Faraday rotation or conversion coefficient.  Altogether there are 11 transfer coefficients.

Following \citet{leung:2011} and \citet{pandya:2016} the polarized synchrotron emissivity coefficients for an isotropic distribution are
\begin{equation}\label{eq:emi_general}
j_{\nu,S}=\begin{pmatrix} j_I \\ j_Q \\j_U\\j_V \end{pmatrix}=\frac{2\pi e^2 \nu^2 n_e}{c} \int d^3 p f(p) \sum_{n=1}^{\infty} \delta(y_n) K_S.
\end{equation}
where $p$ is the momentum and $f(p) \equiv (1/n_e) d n_e/d^3p$
is the distribution function per particle rather than per unit volume; similarly $f(\gamma) \equiv (1/n_e) dn_e/d\gamma$, where $\gamma$ is the electron Lorentz factor. 
Here $S = I,Q,U,V$, $\delta$ is the Dirac delta function and $K_S$ and $y_c$ functions are given in e.g. \citet{pandya:2016}.  Evidently the emissivities depend linearly on the distribution function.  We will work explicit examples for $j_\nu$ later on, but the stochastic averaging procedure can be applied to all transfer coefficients since all depend linearly on $f$, which must in general be derived from kinetic theory.  

One commonly used model for $f$ is the thermal or Maxwell-J\"uttner distribution
\begin{equation}
f_{th}(\gamma)=\frac{\gamma\sqrt{\gamma^2-1}}{\Theta_e
  K_2(1/\Theta_e)} e^{-\gamma/\Theta_e}
\end{equation}
where $\Theta_e=k_B T_e /(m_e c^2)$ is the dimensionless plasma temperature. Notice that if $\Theta_e \gg 1$ then $K_2(1/\Theta_e) \approx 2\Theta_e^2$.  Accurate approximate expressions for the thermal transfer coefficients are available in \cite{marszewski:2021}.

A second commonly used model for $f$ is the $\kappa$ distribution.  A relativistic $\kappa$ distribution proposed by \citet{xiao:2006} reads:
\begin{equation}\label{eq:kappa}
f_{\kappa}(\gamma)=N \gamma \sqrt{\gamma^2-1} \left(1+\frac{\gamma-1}{\kappa w}\right)^{-(\kappa+1)} 
\end{equation}
where $\kappa$ and $w$ are parameters, and $N$ is the normalization constant.  For $\kappa w \gg 1$, $N(\kappa,w) \approx (\kappa-2)(\kappa-1)/2\kappa^2 w^3$.  Evidently $f_\kappa$ smoothly connects a thermal core at low $\gamma$ to a power-law tail at high $\gamma$, which is more convenient than stitching a power-law tail to a Maxwellian core \citep[e.g.,][]{ozel:2000,zhao:2023}.  In the limit $\kappa \rightarrow \infty$, $f_\kappa \rightarrow f_{th}$ with $\Theta_e \equiv k_B T_e/m_e c^2 = w$.  Notice that $N$ goes to zero as $\kappa \rightarrow 2$ from above, i.e. the number of electrons in the unnormalized distribution diverges.  Notice also that the mean energy per electron diverges for $\kappa \le 3$.   

Now consider a distribution function with a continuous parameter $\PAR$, $f(\gamma; \PAR)$ with a known emissivity $j_\nu(\PAR)$.  For example, $f$ might be the thermal distribution with $\PAR = \Theta_e$.  Suppose that we want the emissivity associated with a new distribution that can be written as a stochastic average over $f$:
\begin{equation}
    f_{avg}(\gamma) = \int \, d\PAR \, F(\PAR) \, f(\gamma; \PAR)
\end{equation}
where $\int d\PAR F(\PAR) = 1$, so that $F$ is a probability distribution over $\PAR$.  Then evidently
\begin{equation}
    j_{\nu,avg} = \int \, d\PAR \, F(\PAR) \, j_\nu(\PAR).
\end{equation}
This defines the stochastic averaging procedure.  
The averaging procedure can be applied to the emissivities because they are linear in $f$ (Equation [\ref{eq:emi_general}]).  The remaining transfer coefficients -- absorptivities and rotativities -- for a general distribution function can be obtained from a linear transformation of the susceptibility tensor, and the susceptibility tensor can be written as an integral over $df(\gamma)/d\gamma$ (see e.g., \citealt{pandya:2018}).  The absorptivities and rotativities are therefore also linear in $f$ and can be stochastically averaged.

\section{Synchrotron Emission for $\kappa$ Distributions}\label{sec:3}

The synchrotron transfer coefficients for the $\kappa$ function cannot, to our knowledge, be calculated analytically.  Numerical results can be fit with complicated analytic functions that are restricted, due to computational cost, to specific ranges or values of the $\kappa$ parameter.  Here we develop an efficient method for calculating $\kappa$ transfer coefficients from a stochastic average of thermal distribution functions.  

\subsection{$\kappa$ Distribution from Stochastic Averaging}

Following \citealt{schwadron:2010}, consider a stochastic average over thermal distributions with 
\begin{equation}
    F(\lambda \equiv \Theta_e^{-1}) = \frac{1}{\lambda_0 \Gamma(1-q)}  e^{-\lambda/\lambda_0} \left(\frac{\lambda_0}{\lambda}\right)^q
\end{equation}
where $q < 1$ is a parameter, $\lambda_0$ is a normalizing constant and $\lambda=1/\Theta_e$. Then  
\begin{equation}\label{eq:stochastic_avg}
f_{avg} (\gamma) = \frac{1}{\Gamma(1-q) \lambda_0} \,\,\int_0^{\infty} e^{-\lambda/\lambda_0}
\left(\frac{\lambda_0}{\lambda}\right)^{q}
 \frac{\gamma\sqrt{\gamma^2-1}}{2\frac{1}{\lambda^3}}e^{-\gamma \lambda} d\lambda
\end{equation}
assuming $\Theta_e \gg 1$ so that $K_2(1/\Theta_e) \approx 2 \Theta_e^2$.  

Factoring out the constants:
\begin{equation}
f_{avg} (\gamma)= \frac{1}{2 \Gamma(1-q)} \lambda_0^{q-1} \gamma \sqrt{\gamma^2-1}
\int_0^{\infty} \lambda^{3-q} e^{-\lambda (\gamma+\frac{1}{\lambda_0})} d\lambda
\end{equation}
Next, with the following change of variables: $\lambda'=\lambda
(\gamma+\frac{1}{\lambda_0})$ (and $d\lambda'=d\lambda(\gamma+\frac{1}{\lambda_0})$) and taking 
$q=3-\kappa$ and $\lambda_0=(\kappa w)^{-1}$,
\begin{equation} 
f_{avg} (\gamma)= \frac{1}{2 \Gamma(\kappa-2)} (\frac{1}{\kappa w})^{3} \gamma \sqrt{\gamma^2-1}
\left(\frac{\gamma}{\kappa w}+1 \right)^{-(\kappa+1)}
\int_0^{\infty} \lambda^{'(\kappa)} e^{-\lambda'} d\lambda'
\end{equation}
or
\begin{equation}
f_{avg} (\gamma)= \frac{(\kappa-2)(\kappa-1)}{2 \kappa^2 w^3}  \gamma \sqrt{\gamma^2-1}
\left(\frac{\gamma}{\kappa w}+1 \right)^{-(\kappa+1)}
\end{equation}
Notice that $f_{avg}$ differs from  $f_\kappa$ because the term in parentheses is $\gamma/(\kappa w)$ rather than $(\gamma - 1)/(\kappa w)$.  The difference is negligible, however, if $\kappa w \gg 1$.  This shows that $f_\kappa$ can be approximated with a stochastic average over $f_{th}$.  

So far all we have done is reproduce $f_\kappa$ in approximate form.  Now suppose that we introduce a cutoff in energy by setting $F = 0$ for $\lambda < \lambda_{min}$.  Then 
\begin{equation}
    F(\lambda) = \frac{1}{\lambda_0 \Gamma(1 - q, \lambda'_{min})} e^{-\lambda/\lambda_0} \left(\frac{\lambda_0}{\lambda}\right)^q
\end{equation}
where now $\Gamma$ denotes the upper incomplete $\Gamma$ function and $\lambda'_{min} = \lambda_{min} w \kappa$.  Then
\begin{equation}
    f_{avg} (\gamma) = \frac{\Gamma(1 + \kappa, \lambda'_{min} (1 + \gamma(\kappa w)^{-1} ))}{2 \kappa^3 w^3 \, \Gamma(\kappa - 2, \lambda'_{min})} \gamma \sqrt{\gamma^2 - 1} \left(\frac{\gamma}{\kappa w} + 1\right)^{-(\kappa + 1)} 
\end{equation}
and this immediately yields an approximate expression for the distribution function (which can be expanded in the limit of small $\lambda_{min}'$) as well as a method for obtaining consistent, efficient-to-evaluate transfer coefficients.  

\subsection{Synchrotron emissivities, absorptivities and rotativities using stochastic averaging}

Combining Equation~\ref{eq:emi_general} with~\ref{eq:stochastic_avg} gives an expression for synchrotron emissivity of the approximate $\kappa$ distribution function:
\begin{equation}\label{eq:emi}
j_{\nu,S,avg}(w,\kappa) = \frac{1}{\lambda_0 \Gamma(1-q)} \int_0^{\infty} e^{-\lambda/\lambda_0}
\left(\frac{\lambda_0}{\lambda}\right)^{q}
j_{\nu,S,th}(\lambda) d\lambda
\end{equation}
where $q=3-\kappa$, $\lambda_0=\frac{1}{w\kappa}$ and
$\lambda=\frac{1}{\Theta_e}$ and $j_{\nu}$ is the emissivity of the thermal distribution function.

Similarly for the absorptivities
\begin{equation}\label{eq:abs}
\alpha_{\nu,S,avg}(w,\kappa)= \frac{1}{\lambda_0 \Gamma(1-q)} \int_0^{\infty} e^{-\lambda/\lambda_0}
\left(\frac{\lambda_0}{\lambda}\right)^{q} \alpha_{\nu,S,th}(\lambda) d\lambda.
\end{equation}
and the rotativities 
\begin{equation}\label{eq:rot}
\rho_{\nu,S,avg}(w,\kappa)= \frac{1}{\lambda_0 \Gamma(1-q)} \int_0^{\infty} e^{-\lambda/\lambda_0}
\left(\frac{\lambda_0}{\lambda}\right)^{q} \rho_{\nu,S,th}(\lambda) d\lambda.
\end{equation}
i.e. Faraday rotation ($\rho_V$) and conversion ($\rho_{Q}$, $\rho_U$) coefficients. 

\citet{leung:2011}, \citet{pandya:2016}, \citet{dexter:2016} and \citealt{marszewski:2021} evaluated $j_{\nu,S}$ and $\alpha_{\nu,S}$ or $\rho_{\nu,QV}$ for $f$ thermal, power-law and $\kappa$ functions. Given accurate analytic thermal coefficients $j_{\nu,S}$, $\alpha_{\nu,S}$, and $\rho_{\nu,QV}$, numerical integration of Equations~\ref{eq:emi}, \ref{eq:abs}, and \ref{eq:rot} is computationally cheap and in practice requires a sum over no more than 50 thermal-components.

\begin{figure*}
  \centering
  \includegraphics[width=1.0\linewidth]{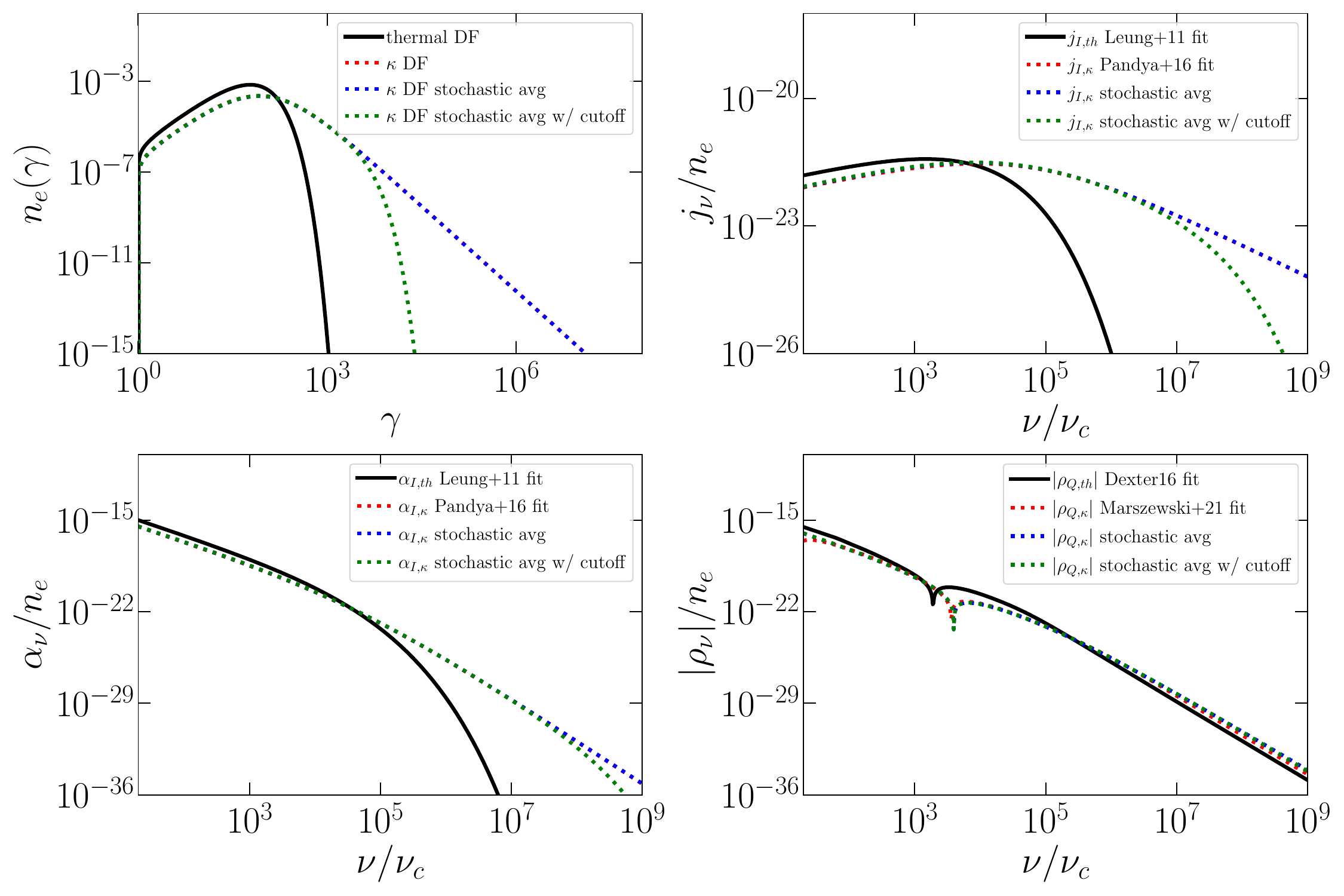}
  \caption{{\bf Top left panel:} electron distribution functions (DF). {\bf Top right and bottom panels:}
    synchrotron emissivities, absorptivities and rotativities for thermal and $\kappa$ distribution functions using fit functions from \citet{leung:2011},\citet{dexter:2016}, \citet{pandya:2016} and \citet{marszewski:2021} 
    in comparison to emissivities, absorptivities and rotativities formulas calculated using stochastic averaging. Purely thermal emissivities are shown for reference. Comparison is shown for the following parameters: $\kappa=3.5$,
    $w=30$, $\theta=60$deg and $B=30$ G (for which $\nu_c\equiv eB/(2\pi m_e c)=0.84 \times 10^8$ Hz). Model with cutoff assumes $\lambda_{min}=0.001$ corresponding
    to averaging end at $\Theta_{e,max}$=1000.}\label{fig:1}
\end{figure*}

The results can be compared to transfer coefficients evaluated numerically by the above authors assuming the $\kappa$ electron distribution function.  Figure~\ref{fig:1} shows examples of thermal, original $\kappa$ and our averaged distribution functions together with their synchrotron emissivities, absorptivities (Stokes I is shown as an example) and rotativities (Stokes Q is shown as an example). The figure shows stochastically averaged emissivities, absorptivities and rotativities in comparison to $\kappa$ fit functions from \citealt{pandya:2016} and \citealt{marszewski:2021}. There is a good agreement between these two. For emissivities/absorptivities the difference is at most 6\%/7\%. Given that we use here only fit functions and our averaged eDF is not exactly the $\kappa$ function, this agreement is remarkable. There is also good agreement between the $\rho_Q$ coefficient obtained from stochastic averaging of thermal coefficient from \citealt{dexter:2016} and $\rho_Q$ for $\kappa$ distribution function from \citealt{marszewski:2021}.

While we find a very good agreement between transfer coefficients for $\kappa$ DF and for thermal DF averaged without the cutoff, calculating transfer coefficients with a cutoff ($\lambda_{min}$=0.001) shows that applying a cutoff to the DF is not equivalent to multiplying the transfer coefficients by the same exponential (or other) frequency cutoff factor \citep[e.g.,][]{davelaar:2019}, which may dominate the errors if the cutoffs are known precisely. The method presented above offers a convenient solution to this limitation.

\subsection{Emission from accelerated electrons near black hole event horizon}

The main application of the presented scheme for computing transfer coefficients is modeling emission (images as well as broadband spectral energy distributions) from non-thermal electrons in relativistic accretion flows and jets from black holes. Given the general transfer coefficients, which are no longer limited to certain values or ranges of the $\kappa$ parameter, we can introduce more realistic (smooth) model for electron acceleration. In this exercise, the underlying model for plasma density, velocity, and magnetic fields is provided by a 3D general relativistic magnetohydrodynamics (GRMHD) simulation of magnetically arrested disk accretion around non-rotating black hole evolved for 30,000 $GM/c^3$ using GRMHD code \texttt{ebhlight} \citep{ryan:2015}. The details of the model setup are comparable to simulations presented in \citet{paper5:2022}. 

The simulations are ray-traced with polarized relativistic radiative transfer code \texttt{ipole} \citep{moscibrodzka:2018}, where the scheme for averaging emissivities, absorptivities and rotativities is implemented. 
When ray-tracing, the GRMHD simulations are scaled to Galactic Center supermassive black hole system, Sgr~A*. It has been long thought that flaring behaviour of this accreting black hole is a manifestation of electron acceleration \citep[e.g.][]{ozel:2000}. Here we present example \texttt{ipole} images of Sgr~A* at single frequency of 230~GHz which roughly corresponds to the peak of the synchrotron emission and the observing frequency of Event Horizon Telescope (EHT). The purpose of this calculation is to illustrate the flexibility and performance of a ray-tracing scheme equipped with stochastically averaged transfer coefficients.

\begin{figure*}
  \centering
  
   \includegraphics[trim={0 12.5cm 0 12.5cm},clip,width=1\linewidth]{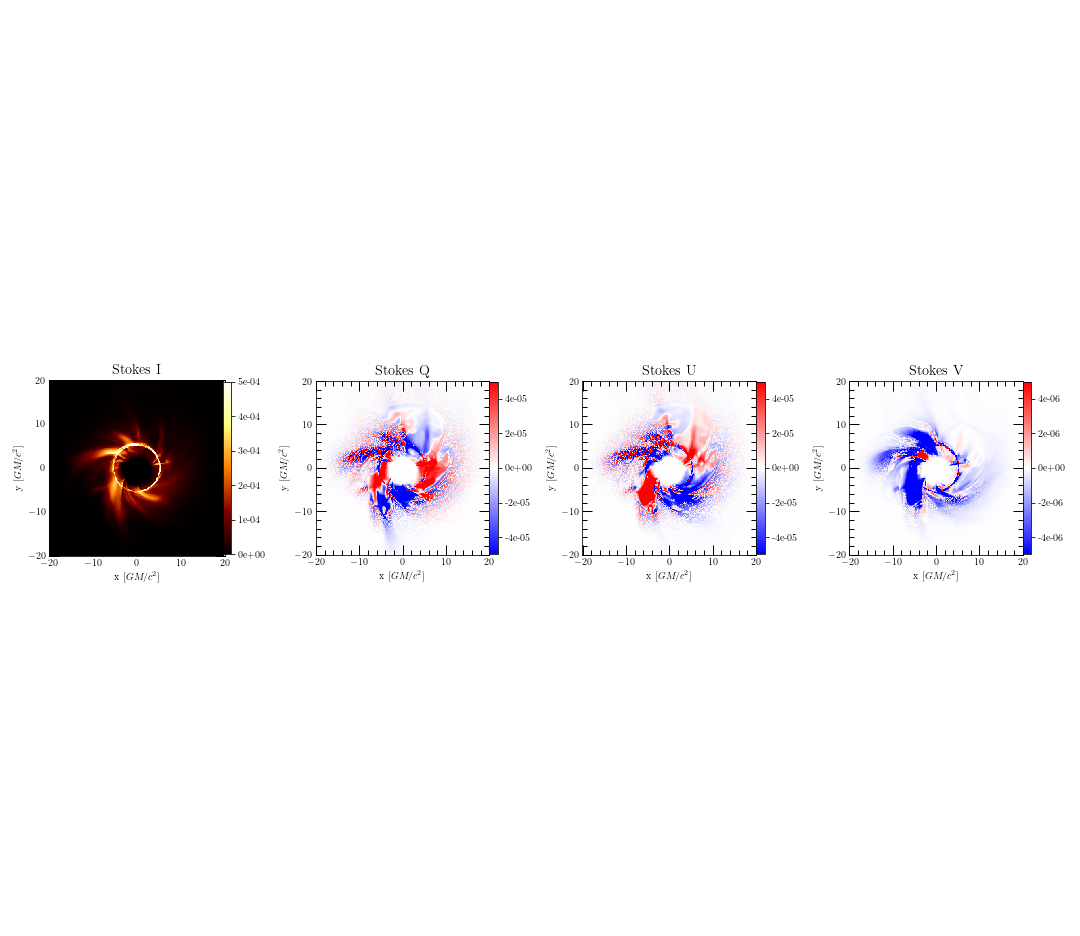}\\
   \includegraphics[trim={0 12.5cm 0 12.5cm},clip,width=1\linewidth]{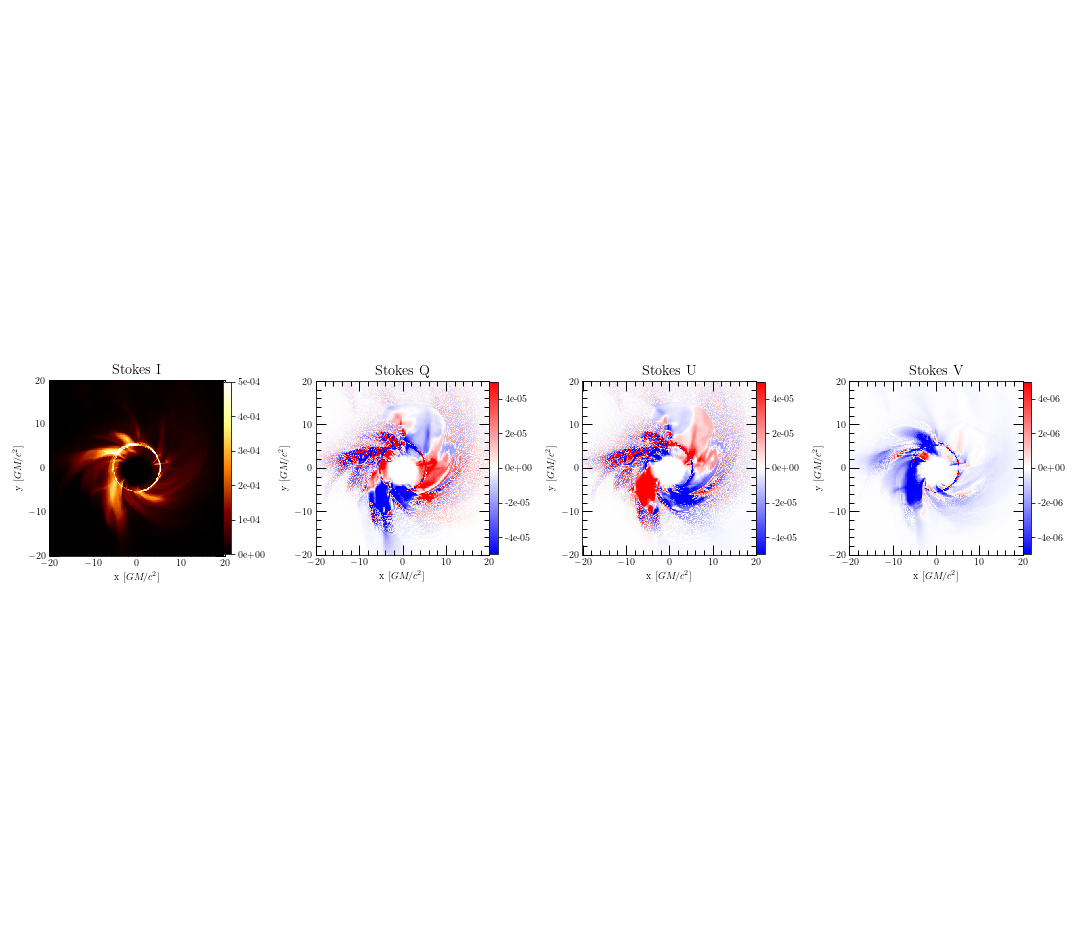}
   \caption{230~GHz polarimetric images of the magnetically arrested accretion flow around non-spinning supermassive black holes produced by ray-tracing simulations. Purely thermal and thermal+non-thermal models are shown in the top and bottom panels, respectively. In the bottom panels, the emission regions are slightly more extended.}\label{fig:2}
\end{figure*}

The radiative transfer calculation makes the following assumptions about the electron distribution functions, which are not followed by the GRMHD simulation. 
The basic model assumes thermal electrons where electron temperatures $\Theta_e$ are given by the $R-\beta$ formula from \citealt{moscibrodzka:2016}, where $\beta \equiv P_{\rm gas}/P_{\rm mag}$ is provided by the underlying GRMHD simulation. The electron temperatures are derived from formula:
\begin{equation}
R=\frac{T_{\rm p}}{T_{\rm e}} = R_{\rm high} \frac{\beta^2}{\beta^2+1} + R_{\rm low} \frac{1}{\beta^2+1}
\end{equation} 
where $R_{\rm high}=160$ and $R_{\rm low}=1$ and proton temperatures, $T_{\rm p}$, are followed in the GRMHD simulation.
Each non-thermal DF is centered around $w=\Theta_e$ where $\Theta_e$ is given by the thermal model but
we allow non-thermal electron distribution function only above some threshold temperature
 $\Theta_{\rm e}>5$.  Apart from this
floor for the electron acceleration we also set a ceiling for acceleration energies by setting $\lambda_{min}=0.001$, corresponding to a 
maximum Lorentz factor of electrons $\gamma \approx 10^4$. 
Finally, we assume that the $\kappa$ parameter is variable and, similarly to $\Theta_e$ and $w$ parameters, is a smooth function of $\beta$:
\begin{equation}\label{eq:kappabeta}
\kappa (\beta) = \kappa_{\rm high} \frac{\beta^2}{\beta^2+1} + \kappa_{\rm low} \frac{1}{\beta^2+1}.
\end{equation}  
where $\kappa_{\rm low}$ and $\kappa_{\rm high}$ are the model free parameters. We set $\kappa_{high}=30$ and $\kappa_{low}=3.5$ for which there is a smooth transition between nearly-thermal electrons in the high-$\beta$ regions (usually closer to the disk equatorial plane) and non-thermal electrons in the low-$\beta$ regions (usually in the jet regions). 
Together these conditions guarantee that $\kappa w \gg 1$ everywhere where a non-thermal DF is allowed.  Notice that equation~\ref{eq:kappabeta} would be difficult to implement with existing fits to the $\kappa$ distribution, which extend only to $\kappa = 7$.

Figure~\ref{fig:2} shows images of the GRMHD simulations assuming: purely thermal electrons (top panels) and non-thermal electrons (bottom panels). Notice that the thermal and non-thermal models have been scaled with slightly different mass scaling unit $\mathcal{M}=8\times 10^{17}$ (non-thermal) and $\mathcal{M}=10\times 10^{17}$ (thermal) so that in both cases the flux density is $\approx$ 2 Jy, characteristic for Sgr~A*.
The differences between purely thermal and non-thermal images at the chosen frequency are only subtle at this observing frequency.\footnote{We have checked that images at this frequency are insensitive to the cutoff, $\lambda_{min}$, which makes a negligible change to the distribution function normalization and removes electrons at $\gamma \gtrsim 10^4$ that do not contribute to emission in the millimeter for parameters appropriate to Sgr A*.} In particular non-thermal Stokes I images are slightly more extended in size compared to the thermal ones. This is consistent with findings of \citet{mao:2017}. Interestingly, the polarimetric (Stokes Q,U,V) images are also slightly more extended. 
With more extended images the ratio of direct emission and lensed (into a photon ring) emission fluxes changes; this may have an impact on the variability of the light curves synthesized from the images.

The integration of the image with stochastically averaged transfer coefficients took only approximately 1.5 times longer compared to the purely thermal image. It is therefore feasible to carry out more extensive parameter surveys of the non-thermal models, including models with different assumptions than those presented here.

\section{Application to Coarse-Graining of Turbulent Emission}\label{sec:4}

The finite resolution of GRMHD models inevitably cuts off turbulent structures close to and below the grid scale $\mil$.  The unresolved structures contain a distribution of temperatures, densities, field strengths, frequency of emission, and field orientations (collectively, the parameter vector $\bPAR$) that may cause the radiative transfer coefficients averaged over a region the size of a cell to differ from the coefficients associated with the average state, since the transfer coefficients are nonlinear functions of $\bPAR$.  Here we apply the notion of stochastic averaging to explore the implications of turbulent sub-structure for the total intensity emissivity alone; similar considerations apply to the other transfer coefficients.

Using stochastic averaging, we define the turbulent emissivity
\begin{equation}
    j_{turb} \equiv \int d^n\bPAR \, F(\bPAR) \, j_\nu(\bPAR)
\end{equation}
where $F$ is the distribution of the $n$ parameters.  

If the variations in emissivity parameters within a zone are small, and we do not need to capture the instantaneous distribution within each zone but only the ensemble-averaged distribution, then we may not do too badly with the ansatz that the parameters $\bPAR$ are distributed like a multivariate Gaussian, i.e. 
\begin{equation}
    F(\bPAR) = ((2 \pi)^n |\Sigma|)^{-1/2} \exp\left(- \frac{1}{2} \Delta \PAR_i \, \Sigma^{-1}_{ij} \,\Delta \PAR_j\right)
\end{equation}
where $\Sigma_{ij} = C_{ij} \sigma_i \sigma_j$ is the covariance matrix, $C_{ij}$ is the dimensionless correlation matrix, and $\Delta \PAR_i \equiv \PAR_i - \langle \PAR_i \rangle$.  

Expanding to second order in $\Delta \PAR$, 
\begin{equation}
    j_{turb} \approx j_\nu \left(
    1 + \frac{1}{2} \frac{1}{j_\nu} \frac{\partial^2 j_\nu}{\partial \PAR_i\partial \PAR_j} C_{ij} \sigma_i \sigma_j
    \right)
\end{equation}
where $j_\nu$ and its derivatives are evaluated at $\langle\bPAR \rangle$, and repeated indices $i,j$ are summed over.  The first order term vanishes because $\langle \Delta \PAR_i \rangle = 0$ by definition.   Notice however that variation of emissivity with $n_e$ enters the turbulent emissivity if density varies in a correlated way with other parameters, as might be the case if density and temperature vary but pressure is approximately constant.  

To go further, we need (1) the standard deviation $\sigma_i$ for each emissivity parameter $\PAR_i$; (2) the correlation matrix $C_{ij}$; and (3) the Hessian of the emissivity with respect to its parameters.

We start by estimating the standard deviation $\sigma_{\PAR}$ for each  parameter $\PAR_i$ inside a zone of size $\mil$.  This is related to the structure function of the turbulence at scale $\mil$, i.e. $\langle (\PAR_i - \langle \PAR_i \rangle)^2 \rangle^{1/2}$. For a passive scalar $\PAR_i$ in turbulence obeying Kolmogorov statistics with outer scale $L$ and variance at the outer scale $\sigma_\PAR(L)$,
\begin{equation}
    \sigma_\PAR(\mil) \sim \sigma_\PAR(L) (\mil/L)^{1/3}.
\end{equation}
For a thermal distribution the emissivity parameters are: electron density $n_e$; electron temperature $\Theta_e$; magnetic field strength $B$; the angle between the line of sight and the magnetic field $\theta$ (the observer angle); and the frequency on emission $\nu$, which is Doppler shifted to the mean frequency of emission in a zone.  

For each parameter we assume the outer scale $L \sim H$, where $H$ is the disk scale height, that the resolution is $N \sim 10^2$ zones per $H$, so that $(\mil/L)^{1/3} \sim N^{-1/3} \sim 0.2$.   For $n_e$, assuming $\sigma_{n_e}(L) \sim n_e$, $\sigma_{n_e}(\mil) \sim 0.2 n_e$.  For $\Theta_e$, $\sigma_{\Theta_e}(L) \sim \Theta_e$, $\sigma_{\Theta_e}(\mil) \sim 0.2 \Theta_e$.  For $B$, $\sigma_B(L) \sim B$, $\sigma_B(\mil) \sim 0.2 B$.  For $\theta$, $\sigma_\theta(L) \sim 1$, $\sigma_\theta(\mil) \sim 0.2$ rad.  For frequency $\nu$, we assume the velocity structure function $\sigma_v(L) \sim c_s$ and thus $\sigma_v(\mil) \sim 0.2 c_s$, so the frequency of emission, in the plasma frame, fluctuates by $\sigma_\mil(\nu)/\nu \sim (0.2/\sqrt{3}) c_s/c \simeq 0.12 c_s/c$, where the factor of $\sqrt{3}$ makes the velocity dispersion one-dimensional.  

Next consider the correlation matrix $C_{ij} \equiv \Sigma_{ij}/(\sigma_i \sigma_j)$.  This is a property of turbulence and thus of the flow.  It will vary with the global flow structure and also with lengthscale.

For definiteness we evaluate two examples using a set of GRMHD simulations of aligned accretion around a black hole with spin $a = 0.75$.  The simulations have adiabatic index $\gamma = 5/3$ and resolution $384 \times 192 \times 192$.  We compute the covariance in a narrow annulus within $0.1\pi$ rad of the midplane, with $3.8 < r c^2/(G M) < 4.2$.  For simplicity we ignore the velocity (frequency) variations and the variations in field direction, and consider only the gas temperature $\Theta = P/(\rho c^2)$ rather than using a model for the electron temperature, which - because electron temperature assignment models commonly depend on plasma $\beta$ - can alter the correlations.  Thus we consider correlations between $\rho$, $\Theta$, and $B$, averaged over azimuth and time.  

The diagonal elements of the correlation matrix are $1$ and it is symmetric, so there are 3 nontrivial entries.  For a SANE model these are 
$C_{\rho\Theta} = -0.12$, 
$C_{\rho B} = -0.38$, and
$C_{\Theta B} = -0.20$.  For a MAD model these are 
$C_{\rho \Theta} = -0.08$, 
$C_{\rho B} = -0.48$, and
$C_{\Theta B} = -0.09$.  Interestingly these variations show that MAD and SANE model correlations differ.  Both show strong anticorrelation between $\rho$ and $B$.  

Finally consider the Hessian of $j_{\nu,I}$ with respect to its parameters. To make an estimate of the importance of turbulence in changing the emissivity we must choose an emissivity and therefore a distribution function.  We use a thermal distribution.  Defining 
\begin{equation}
    A \equiv \frac{n_e e^3 B \sin \theta}{27 \sqrt{2} m_e c^2}; \qquad \qquad X \equiv \frac{9 \pi m_e c \nu}{\Theta_e^2 \, e B \sin \theta};
\end{equation}
and working in the limit $X \gg 1$, 
\begin{equation}\label{eq:thermj}
    j_\nu \simeq A \, X \, \exp^{-X^{1/3}}.
\end{equation}
The general expression for the Hessian is readily evaluated but not very interesting.  Again in the interests of getting to an estimate we adopt values for the parameters appropriate to the EHT  source Sgr A*:
\begin{equation}
    n_e = 10^6\, \mathrm{cm}^{-3}, \quad B = 30\, \mathrm{G}, \quad \Theta_e = 10, \quad \theta = \pi/3, \quad \nu = 230\mathrm{GHz}
\end{equation}
and consider only variations in the parameters $n_e$, $\Theta_e$, and $B$, assuming that $\Theta_e$ varies like $\Theta$.  Notice that for these parameters $X \sim 10^2$, justifying our use of Equation (\ref{eq:thermj}).  Then at last we find for the MAD models
\begin{equation}\label{eq:turbcorr}
    \frac{1}{2} \frac{1}{j_\nu} \frac{\partial^2 j_\nu}{\partial \PAR_i\partial \PAR_j} C_{ij} \sigma_i \sigma_j \approx -0.06
\end{equation}
That is, turbulence reduces the volume-averaged emissivity by $6\%$.  The largest contribution to the sum in (\ref{eq:turbcorr}) is from the $(i,j) = (\Theta_e, \Theta_e)$ term, and is negative because the emissivity is concave with respect to $\Theta_e$.  The second largest term is from the $(n_e, B)$ terms, and arises only because these quantities are anticorrelated.  To sum up: for the EHT-relevant parameters considered here the emissivity correction due to turbulence is small and negative. 

Small negative corrections due to turbulence will alter the Sgr A* model normalization: a higher accretion rate, and therefore stronger magnetic fields, are needed to achieve the same millimeter flux density. This may be a point of concern for EHT modelers in both Sgr A* and M87*, since it will shift the accretion rate and jet power up, and potentially change the polarization properties.  

Finally, notice that there are circumstances in which the turbulent correction is large. Consider the correction in the near infrared where the emissivity is dropping exponentially with increasing frequency.  Then the {\it dominant} term in Equation~\ref{eq:turbcorr} is the $(\Theta_e,\Theta_e)$ term, which is $\sim 14$.  The second largest is the $(B,B)$ term, which is $\sim 4$.  Evidently if the Hessian is large then the emergent radiation can be very sensitive to turbulent corrections.

\section{Concluding remarks}\label{sec:5}

We have developed a notion of stochastic averaging of radiative transfer coefficients.  This procedure, which is just a weighted sum of transfer coefficients over a distribution of parameters, can provide a quick, practical method for evaluating transfer coefficients.   

In particular we have shown that one can apply stochastic averaging to construct radiative transfer coefficients for the $\kappa$ distribution function from those for a thermal distribution. The calculation presented here is inaccurate for mildly relativistic plasmas,
though mildly relativistic regions may not contribute significant emission anyway. 

Stochastic averaging of transfer coefficients is valid for a non-relativistic electron distribution function, as shown by \citet{schwadron:2010}; the non-relativistic electrons distributed into $\kappa$ function will not emit synchrotron photons but can still contribute to Faraday rotation.  

We have also applied stochastic averaging to estimate corrections to the emitted radiation from subgrid scale turbulence.  We find that for parameters appropriate to Sgr A* in the millimeter, the turbulent corrections are small and negative.  In the near-infrared the turbulent corrections are large (an order of magnitude) and positive.  
The turbulent emissivity calculation is  valid for relativistic, mildly-relativistic and sub-relativistic plasmas.

The applications presented in this work stochastically average over a thermal distribution, and are therefore trivially applicable only to isotropic distributions.  It would be interesting to generalize to anisotropic distributions \citep[e.g.][]{galishnikova:2023}.

As a next step, one might also incorporate the stochastic averaging scheme into Compton scattering kernels to predict high energy emission from scattering of synchrotron photons on non-thermal electrons or scattering on turbulent substructures which are unresolved in the GRMHD models. 
Scattering off turbulent Compton kernel can be thought of as a parametric variability model which can be then tested observationally e.g. by X-ray observations. 

Finally, we emphasize that the covariance between emissivity parameters needed in calculating the turbulent emissivity is a comparatively unstudied characteristic of turbulence (as of this writing we are not aware of other calculations of the covariance for MHD turbulence).  It would be interesting to understand how this covariance depends on lengthscale and on flow parameters.

\begin{acknowledgments}

The work presented in this paper was initiated at workshop Modeling Plasmas Around Black Holes
hosted by Lorentz Center in Leiden in fall of 2023.
MM acknowledges support from Dutch Research Council (NWO), grant
no. OCENW.KLEIN.113 and NWO Athena Award. Regarding calculations in section 3.3, authors gratefully acknowledge the HPC RIVR consortium (www.hpc-rivr.si) and EuroHPC JU (eurohpc-ju.europa.eu) for funding
this research by providing computing resources of the HPC system Vega at the Institute of Information Science (www.izum.si).
We are grateful to Ben Prather and Vedant Dhruv for providing some of the calculations used in Section 4. This work was supported by NSF AST 20-34306.  This research used resources of the Oak Ridge Leadership Computing Facility at the Oak Ridge National Laboratory, which is supported by the Office of Science of the U.S. Department of Energy under Contract No. DE-AC05-00OR22725.  This research used resources of the Argonne Leadership Computing Facility, which is a DOE Office of Science User Facility supported under Contract DE-AC02-06CH11357.  CFG was supported in part by the IBM Einstein Fellow Fund at the Institute for Advanced Study, and also in part by grant NSF PHY-2309135 and the Gordon and Betty Moore Foundation Grant No. 2919.02 to the Kavli Institute for Theoretical Physics (KITP). We thank the referee for helpful comments that improved the quality of this paper.

\end{acknowledgments}


\end{document}